\documentstyle[12pt,epsfig,amsfonts,amssymb]{article}
\begin{document}
\newcommand{\smU}{{\scriptscriptstyle U}}
\newcommand{\sm}[1]{{\scriptscriptstyle #1}}
\def\d{\partial}
\def\l{\left(}
\def\r{\right)}
\newcommand{\e}{\mathop{\rm e}\nolimits}
\newcommand{\Tr}{\mathop{\rm Tr}\nolimits}
\renewcommand{\Im}{\mathop{\rm Im}\nolimits}
\renewcommand{\Re}{\mathop{\rm Re}\nolimits}
\newcommand{\be}{\begin{equation}}
\newcommand{\ee}{\end{equation}}
\newcommand{\ba}{\begin{eqnarray}}
\newcommand{\ea}{\end{eqnarray}}

%\begin{document}

%%%%%%%%%%%%% Title
\begin{center}
  {\Large\bf On  models of gauge field localization on a
brane} \\
  \medskip S.~L.~Dubovsky, V.~A.~Rubakov

   \medskip
  {\small
     Institute for Nuclear Research of
         the Russian Academy of Sciences,\\  60th October Anniversary
  Prospect, 7a, 117312 Moscow, Russia
  }
  
\end{center}

\begin{abstract}
We argue that any viable mechanism of gauge field localization 
should automatically imply charge universality on the brane. 
We study whether this condition is satisfied in the two known proposals aimed
to localize vector field in flat bulk space. We construct a simple
calculable model with confinement in the bulk and deconfinement on the
brane, as in the Shifman--Dvali set up. We find that in our model
the 4-dimensional Coulomb law is indeed reproduced on the brane
due to the massless localized photon mode. The charge
universality is enforced by the presence of 
``confining strings''. 
On the other hand, charge universality condition is
not satisfied in another, brane-induced localization mechanism when the number
of extra dimensions $d$ is larger than two. We demonstrate that in the
non-Abelian case the gauge fields inside the brane
are never four-dimensional and their self-interaction is 
strong at all distances of interest. Hence this mechanism 
does not work for $d>2$. 
At $d=2$ the charge
universality is still a problem, but it holds automatically at $d=1$. 
At $d=1$, however, the bulk gauge fields are strongly coupled in the
non-Abelian case.
\end{abstract}
\section{Introduction and Summary}
Phenomenological discussions of various ``brane world'' scenarios are
sometimes based on field theoretic models for localization of matter
fields and possibly gravity on a three-dimensional submanifold ---
brane --- embedded in higher dimensional space. A simple mechanism of
the localization of fermions (and also scalars) makes
use of domain walls or topological defects of higher codimension
\cite{Rubakov:1983bb,Akama:1982jy}: in the presence of a topological
defect, there often exist (chiral) fermion zero modes whose wave
functions concentrate near the defect 
\cite{Jackiw:1976fn,Jackiw:1981ee,'tHooft:1976fv}. Somewhat
similar mechanism localizes gravitons in the vicinity of a gravitating
domain wall embedded in five-dimensional space
\cite{Visser:1985qm,Randall:1999vf};
higher-dimensional generalizations of the latter mechanism have been
proposed in 
Refs.~\cite{Gherghetta:2000qi,Gherghetta:2000jf,Randjbar-Daemi:2000ft} 
(see,
however, Ref.~\cite{Ponton:2001gi}). Gravity of a brane is capable of
localizing scalars~\cite{Bajc:2000mh} and gauge
fields~\cite{Oda:2000zc,Dubovsky:2000av} as well (for a review see, e.g.,
Ref.~\cite{VR}).

Some brane world scenarios, notably ADD~\cite{Arkani-Hamed:1998rs},
require non-gravitational mechanisms of matter localization. In this
regard, it is of interest to understand possible ways of trapping
{\it gauge fields} to a brane in flat multi-dimensional
space. The simple zero mode mechanism, similar to that localizing
fermions and scalars on a defect, is unlikely to work for gauge
fields, for the following reason. Suppose that $(3+1)$-dimensional
fermions and gauge fields are zero modes of bulk spinors and vectors,
respectively, {\it i.e.} they are described by the wave functions
\begin{eqnarray}
\Psi\l x^{\mu},z^{\alpha}\r&=&\psi\l x^{\mu}\r\Psi_0\l
z^{\alpha}\r\nonumber\\
A_{\mu}\l x^{\nu},z^{\alpha}\r&=&A_{\mu}\l x^{\nu}\r{\cal A}\l
z^{\alpha}\r\;,\nonumber
\end{eqnarray}
where $x^{\mu}$ ($\mu=0,1,2,3$) and $z^{\alpha}$ ($\alpha=1,\dots,d$) are
coordinates along the brane and transverse to the brane, respectively,
$\psi\l x^{\mu}\r$ and $A_{\mu}\l x^{\nu}\r$ are the usual
four-dimesional wave functions (say, plane waves) and $\psi_0\l
z^{\alpha}\r$ and ${\cal A}\l
z^{\alpha}\r$ are the wave functions in the transverse dimensions. The
latter are supposed to correspond to bound states (massless from the
four-dimensional point of view), so that $\psi_0\l
z^{\alpha}\r$ and ${\cal A}\l
z^{\alpha}\r$ are peaked near the brane, $z^{\alpha}=0$, and decrease
towards $|z|\to\infty$. Multi-dimensional gauge interactions would
then induce interactions between fermions and vectors residing on the
brane, with the fermion gauge charges in the four-dimensional effective
theory being proportional to the overlap integrals
\be
\label{5*}
\int d^dz\Psi_0^{\dagger}\l z^{\alpha}\r {\cal A}\l
z^{\alpha}\r\Psi_0\l z^{\alpha}\r\;.
\ee
The problem is that these overlap integrals may take arbitrary values:
the shapes of the fermion zero mode wave functions depend on the
details of the interaction of bulk fermions to the defect and are
different, at least in principle, for different fermionic
species. Hence, the zero mode mechanism would allow for different (and
arbitrary) values of gauge charges for different types of fermions,
which should not be possible, at least in the non-Abelian case. 

Any viable mechanism of gauge field localization should automatically
lead to equal gauge charges of matter fields residing on the brane,
{\it i.e.}, should automatically ensure charge universality. This
property is not inherent in the simple zero mode mechanism in flat
background, so it has no chance to work\footnote{It is worth
commenting on
how the charge universality obstruction is avoided in models
where gauge fields are localized by
gravity~\cite{Oda:2000zc,Dubovsky:2000av}. In these models, the gauge
field zero mode ${\cal A}$ is {\it independent} of $z$, and yet
it is normalizable with appropriate measure determined by the bulk
geometry. The overlap integral analogous to (\ref{5*}) coincides then
with the norm of $\Psi_0$, so gauge charges are in fact universal.}.

In this paper we discuss two proposals for localizing gauge
fields. One of them~\cite{Dvali:1997xe} is based on the assumption
that the gauge theory is confining in the bulk, whereas confinement is
absent on the brane\footnote{We note in passing that, to the best of our
knowledge, a "microscopic"
higher-dimensional quantum field theory possessing 
these properties, is yet unknown.}. In Section 2 we substantiate 
this proposal by
modelling this situation in a theory of dual
superconductivity. For simplicity, we consider the case of one extra
dimension; then this theory contains a two-form field with the mass
parameter $\Lambda(z)$ depending on the extra coordinate $z$. For
constant $\Lambda$ this theory 
\cite{Quevedo:1997uu,polyakov} possesses the
't Hooft--Mandelstam mechanism of
confinement. We will see that with $\Lambda(z)$ vanishing on the
brane, charges residing on the brane experience four dimensional
Coulomb law, and that the zero mode of the gauge field appears. We
will also discuss how charge universality is enforced by confinement
in the bulk.

Another proposal~\cite{Dvali:2001rx} is to modify the action of the
gauge field by adding to conventional bulk action a term
concentrating on a brane; this proposal 
is based on earlier discussion of the
possibility to localize gravity on a brane in this way
\cite{Dvali:2000hr,Dvali:2001xg}. In the limit of zero brane
thickness, the additional term is
\be
S_{brane}\propto \int d^4x~d^dz~\delta^d(z)~F_{\mu\nu}^2(x,z)\;.
\label{i*}
\ee
It has been argued that this term may be induced by loops involving
particles residing on the brane, and that there is a wide range of
distances at which 
the gauge theory on the brane is effectively 
four-dimensional~\cite{Dvali:2001rx}.

The brane-induced localization mechanism works differently for one
extra dimension, $d=1$, and larger number of extra dimensions.
At $d=1$, the relevant gauge field propagator is non-singular
on the brane, whereas at larger $d$ it has a singularity at 
$z=0$ (if the $\delta$-function in eq.~(\ref{i*}) is not
regularized). An interpretation of the latter is that for $d>1$,
charges placed exactly on the brane and charges slightly
displaced from the brane interact in a quite different way.
This signalizes for the lack of charge universality, so
one suspects that the brane-induced localization may have problems at
$d>1$. 

In fact, the regularization of the $\delta$-function in the action
(\ref{i*}) is a necessity, otherwise there are uncontrollable
singularities. The simplest regularization is smearing this
$\delta$-function out (cf. Ref.~\cite{Dvali:2001rx}, another
regularization is proposed in Ref.~\cite{Carena:2001xy}).
In Section 3 we adopt this approach and consider the case $d>2$.
We find that, in a sense, the gauge field inside the brane is never
four-dimensional, which is unacceptable in non-Abelian
theory. Furthermore, in the non-Abelian case the gauge coupling in the
theory inside the brane is strong at all distances of interest. We
conclude that this model does not admit semiclassical
treatment, so the mechanism of Ref.~\cite{Dvali:2001rx}, as it stands,
does not work for non-Abelian gauge fields and $d>2$.

While at $d=2$ the charge universality is still a problem,
the situation is different at $d=1$.
This case is also considered in Section 3. 
We find that in this case, the charge universality holds automatically
in effective four-dimensional theory. In non-Abelian gauge theory,
however, there is another problem: if one requires the effective
four-dimensional gauge coupling to be roughtly of order one, the
original five-dimensional theory is strongly coupled in the bulk at
distance scales of interest.

To conclude this section, let us point out that a 
gravitational analogue of the charge universality 
 is the equivalence principle. So, it is natural to 
conjecture that any self-consistent theory 
(quasi-)localizing graviton on a 
brane should automatically lead to the four-dimensional
equivalence principle. This property is indeed present in
the Randall--Sundrum proposal \cite{Randall:1999vf}, but does not 
appear to be inherent in the brane-induced localization mechanism
of Refs.~\cite{Dvali:2000hr,Dvali:2001xg}. We point out in
Section 3 that gravity ``localized'' by the brane-induced mechanism
has the same problems at $d>2$ as gauge fields, whereas at $d=1$ the
four-dimensional equivalence principle is ensured. Hence, if it was
not for the problem with the tensor structure of the graviton
propagator \cite{Dvali:2000hr} and the mismatch of scales
(see Ref.~\cite{Carena:2001xy} and subsection 3.2), the brane-induced
mechanism would be a viable scenario of the localization of gravity
in {\it five} flat dimensions.

\section{Confinement in the bulk, no confinement on the brane}
\subsection{Model for dual superconductivity}
Let us consider space-time of $D$ ``our'' dimensions ($D\geq 3$) and one extra
dimension.
To construct an explicit model with confinement in the bulk and no
confinement on the brane, we begin with the
 theory~\cite{Quevedo:1997uu} exhibiting the 
't Hooft--Mandelstam
mechanism of confinement. 
For the time being,we consider homogeneous $(D+1)$-dimensional
space-time; we will introduce the brane later on.
Microscopically, one
thinks of an Abelian theory with monopoles, and assumes that there is a
phase with monopole condensate. Then electric charges are
confined. Phenomenologically, this situation may be described by the
low energy effective theory involving a two-form field $\omega_{ab}$
($a,b=1,\dots,D+1$) which generalizes the Maxwell field strength. 
The sources are represented by an anti-symmetric
tensor $T_{ab}$ related to the usual electromagnetic current as
follows
\begin{equation}
\label{current}
\d_{b}T^{ab}={1\over 2}j^a\;.
\end{equation}
The action for this model is
\begin{equation}
\label{action}
S=\int d^{D+1}X \l{1\over 12\Lambda^2}\Omega_{abc}\Omega^{abc}+{1\over
4e^2}\omega_{ab}\omega^{ab}+
i\omega_{ab}T^{ab}\r\;,
\end{equation}
where
$\Omega_{abc}$
is the field strength,
\be
\Omega_{abc}=\d_a\omega_{bc}+\d_c\omega_{ab}+\d_b\omega_{ca}\;.
\ee
We work in Euclidean $(D+1)$-dimensional space-time, hence the factor
$i$ in the action (\ref{action}). 
The parameter $\Lambda$ may be interpreted as the energy scale of the
monopole condensate, and $e$ is the electric charge. It is implicit in
(\ref{action}) that the theory has a fixed ultraviolet cut-off $\mu$;
we will see shortly that the tension of the string between electric
charges depends on this cut-off.

As explained in Ref.~\cite{Quevedo:1997uu}, this action can be
obtained by a certain gauge fixing from a more general Abelian action.
At $D=3$ the latter action is dual to
the Abelian Higgs model with frozen vacuum expectation value of the
Higgs field. The two-form field
$\omega_{ab}$ is dual to the phase of the Higgs field.
At $D>3$ the action before gauge fixing is dual to
the generalized Higgs model describing $(D-2)$-form field dual to the
electromagnetic field, and $(D-3)$-form field dual to the
antysimmetric tensor $\omega_{ab}$. This $(D-3)$-form is a generalization
of the phase of Higgs field in the four-dimensional Abelian Higgs model.

It is straightforward to see that when the monopole condensate
vanishes, $\Lambda\to 0$, the model (\ref{action}) reduces to the usual
QED. Indeed, finiteness of the action implies that in this limit 
\be
\Omega_{abc}=0
\ee
and, consequently, 
\be
\omega_{ab}=\d_a A_b-\d_b A_a
\ee
with some vector field $A_a$. Then the relation (\ref{current}) implies
that the last two terms in Eq. (\ref{action})
describe photon field $A_a$ interacting with electromagnetic current
$j_a$. 

For non-vanishing $\Lambda$ the situation is not so simple. In the 
case of two opposite
point charges, Eq.~(\ref{current}) implies that
\be
T^{ab}=-{1\over 2}\int d^2\sigma \epsilon^{\alpha\beta}{\d
x^a\over \d \sigma^{\alpha}}{\d
x^b\over \d \sigma^{\beta}}\delta^{D+1}(x-x(\sigma))
\ee
where $x(\sigma)$ parametrizes a surface $\Sigma$ bounded by the
world-lines of the charges. Hence the action
(\ref{action}) as it stands depends on the choice of this surface. The
potential between electric charges in dual supeconductor should, however,
 depend only on the
locations of these charges. The right way to ensure the latter property is to
integrate over all surfaces $\Sigma$ \cite{polyakov}. This integration
is interpreted as the integration over world sheets of confining
strings.
We will perform this
integration semiclassicaly, taking the surface which minimizes the
action. In most cases considered below this surface is
uniqely determined by symmetries.

The solution to the classical field equations following from the
action (\ref{action}) is
\begin{equation}
\label{flat_field}
\omega_{ab}=-i{2\Lambda^2\over m^2-\Delta_{D+1}}T_{ab}-
i{e^2\over m^2-\Delta_{D+1}}(\d_a j_b-\d_b j_a)\;,
\end{equation}
where $\Delta_{D+1}$ is $(D+1)$-dimensional Euclidian Laplacian, and
mass $m$ is equal to
\[
m={\Lambda\over e}\;.
\]
It is straighforward to check that an analogue of the first pair
of Maxwell's equations is satisfied,
\begin{equation}
\label{conservation}
\d_a\omega_{ab}=ie^2j_b\;.
\end{equation}
This is in accord with the interpretation of
 $\omega_{ab}$ as the generalization of the
Maxwell tensor in the phase with monopole
condensate. Substituting the field (\ref{flat_field}) back into the
action (\ref{action}) and integrating by parts 
 one obtains the following expression for the
action describing the interaction of electric charges,
\begin{equation}
\label{energy}
S=\int d^{D+1}x \l T_{ab}{\Lambda^2\over
m^2-\Delta_{D+1}}T_{ab}+{e^2\over 2} j_a{1\over m^2-\Delta_{D+1}}j_a\r\;.
\end{equation}
The second term in Eq.~(\ref{energy}) is local and does not depend on
the choice of the string world sheet. It reduces to the usual Coulomb
term at small charge separation,
$L\ll m^{-1}$ and exponentially decreases at larger distances. The first
term in Eq.~(\ref{energy}) is non-local and gives rise to the
confining potential between charges.

To see the property of confinement explicitly, let us consider 
two static charges located at the origin and at the point
$X_1=L$, $X_2=\dots=X_D=0$. In this case the minimal surface is clearly
a flat rectangular, infinite in time direction. The
non-vanishing components of $T_{ab}$ are
\begin{equation}
\label{wilson_loop}
T_{01}=-T_{10}=\delta^{D-1}(X_l)\theta(X_1)\theta(L-X_1)
\end{equation}
where $l$ runs from 2 to $D$. In this static case, the action
(\ref{energy}) determines the potential between the charges,
\[
S=V(L)T\;.
\]
The non-local term in Eq.~(\ref{wilson_loop}) gives the following
contribution to the potential (to the leading order in $m/\mu$)
\begin{equation}
\label{Pauli}
V_{conf}(L)=\sigma L\;,
\end{equation}
 where the string tension $\sigma$ explicitly depends\footnote{
Explicit dependence of the string tension $\sigma$ on the cutoff
scale $\mu$ is due to the
divergence of the integral over the coordinates transverse to the
string worldsheet.  Equation~(\ref{sigma}) is in  agreement
with the well-known logarithmic divergence in the energy of the usual
($D=3$) Abrikosov--Nielsen--Olesen vortex in the limit of infinite
Higgs mass.} on the cutoff
 $\mu$,
\begin{equation}
\label{sigma}
\sigma\propto \Lambda^2\mu^{D-3}\;.
\end{equation}
The Coulomb and confinement regimes occur at short and long distances,
the transition between the two takes place at the scale where
confining and Coulomb potentials are of the same order,
\be
\label{Lc}
 L_c\sim \l{e^2\over \sigma}\r^{1\over D-1}\sim  
\l{e^2\over \Lambda^2\mu^{D-3}}\r^{1\over D-1}\;.
\ee
Note that $L_c\ll m^{-1}$ so that in the whole region where the local
contribution dominates, the potential between the charges
is indeed of the Coulomb type.
\subsection{Potential between charges on the brane}
To consider the situation  proposed by Dvali and Shifman
as a
mechanism of the
localization of gauge fields \cite{Dvali:1997xe}, we
modify the above model and take the scale of confinement $\Lambda$ to
be a
non-trivial function in transverse space, which 
is supposed to
vanish on the brane
surface. For the sake of simplicity we consider flat co-dimension
one brane, {\em i.e.}  assume that $\Lambda=\Lambda(z)$ is a function
of one coordinate $z\equiv X^{D+1}$. The brane is placed at $z=0$,
and symmetry $z \to -z$ is assumed.
In
order to avoid singularities in the action (\ref{action}) we work
with small, but non-vanishing $\Lambda(0)$. The corresponding mass
$m_0=\Lambda(0)/e$ is the smallest energy scale in
the problem. We will not specify the explicit form of the function
$\Lambda(z)$ for the moment and require only that outside the region
of small size $z_0$, the confinement scale $\Lambda(z)$ is
constant\footnote{Our analysis remains valid for $\Lambda(z)$ growing
towards $|z|\to\infty$. In fact, this is  the case in a specific
example to be considered later, see Eq.~(\ref{choice}).}
and large enough, $\Lambda(|z|>z_0)\equiv\Lambda_c\gg\Lambda(0)$.

The field equation following from the action (\ref{action}) with 
varying $\Lambda$ has the form
\begin{equation}
\label{eqofmotion}
\d_{a}\l 1/\Lambda^2(z)\Omega_{abc}\r - {1\over e^2}\omega_{bc}-2iT_{bc}=0
\end{equation}
Taking the divergence of this equation we again obtain the
first pair of Maxwell's equations, Eq.~(\ref{conservation}). 
It is straightforward to solve
Eq. (\ref{eqofmotion}) for a general source $T_{ab}$. The result is
\be
\label{wzmu}
\omega_{z\mu}={-i\over{\cal D}_T-\Delta_D}\left[\Lambda^2(z)T_{z\mu}+e^2
 (\d_z j_{\mu}-\d_{\mu}j_z)\right]
\ee
and
\begin{eqnarray}
\label{wmunu}
\omega_{\mu\nu}={-i\over {\cal D}_L-\Delta_D}
\left[\Lambda^2(z)T_{\mu\nu}+e^2
 (\d_{\mu} j_{\nu}-\d_{\nu}j_{\mu})-2i{\Lambda'(z)\over
\Lambda(z)}(\d_{\mu}\omega_{z\nu}-\d_{\nu}\omega_{z\mu}) \right]
\end{eqnarray}
where transverse and longitudial 
operators ${\cal D}_T$ and ${\cal D}_L$ are 
\be
\label{tranD}
{\cal D}_T=m^2(z)-\d_z^2
\ee
and
\be
\label{longD}
{\cal D}_L=m^2(z)-\d_z^2+2{\Lambda'(z)\over\Lambda(z)}\d_z
\ee
Indices $\mu$, $\nu$ run from 0 to $D$, and prime denotes
differentiation with respect to $z$.

Let us consider two static point-like charges located
on the brane at distance $L$ from each other. According to the
qualitative arguments due to Dvali and Shifman, the $D$-dimensional
(rather than $(D+1)$-dimensional) Coulomb potential should emerge
at large distances
in this case.  Since we keep $\Lambda(0)$ finite, charges on the brane
experience confinement at very large distances, so this picture does
not hold beyond the confinement scale  on the brane
\be
\label{confscale}
L_c(0)= \l{e^2\over \Lambda^2(0)\mu^{D-3}}\r^{1\over D-1}\;.
\ee
We will consider the distances $L\ll L_c(0)$, but still larger than
all other scales inherent in the model. Our purpose is to see whether
the $D$-dimensional Coulomb potential between charges on the brane
indeed emerges at these distances.

From the symmetry
$z\to -z$ and from the fact that the confinement scale $\Lambda(z)$ is
minimal at $z=0$ it is clear that the minimal surface in our case is
the same as for  non-vanishing constant $\Lambda$,
so the source $T_{ab}$ is
again given by Eq.~(\ref{wilson_loop}). Then, as in the usual
electrodynamics, only electric field is non-vanishing. From
Eqs.~(\ref{wzmu}) and (\ref{wmunu}) we find
\be
\label{Ez}
E_z\equiv{1\over i}\omega_{z0}=-e^2({\cal D}_T-\Delta_D)^{-1}\d_z j_0
\ee
for the component of the electric field transverse to the brane, and
\begin{eqnarray}
\label{El}
E_{i}\equiv{1\over i}\omega_{i0}=-\l{\cal D}_L-\Delta_D\r^{-1}
\left[\Lambda^2(0)T_{i0}+e^2
 \d_{i} j_{0}+2{\Lambda'(z)\over
\Lambda(z)}\d_{i}E_z \right],
\end{eqnarray}
for the electric field paralel to the brane 
($i=1,\dots (D-1)$). Equation (\ref{conservation})
becomes the $(D+1)$-dimensional Gauss' law,
\be
\partial_z E_z + \partial_i E_i = e^2 j_0
\label{8*}
\ee

The  operator (\ref{tranD}) entering the expression
(\ref{Ez}) for the transverse electric field 
is a Schr\"odinger operator with the potential
$m^2(z)$ having the shape of a well of charteristic width $z_0$ and 
characteristic height $(\Lambda_c/e)^2$. The mass of the lightest mode
corresponding to this operator is of order
\be
m_T\sim\mbox{min}\{z_0^{-1},\Lambda_c/e\}\;
\ee
For small enough $\Lambda(0)$, this mass is much larger than
$L_c(0)^{-1}$. Then a range of intermediate distances
$L$ exists, where $E_z$ component of the electric field is already
exponentially small but confinement on the brane has not yet set
in. We are interested precisely in this range of distances between the
charges.

At these distances the stringy term, $\Lambda^2(0)T_{i0}$, in the
expression (\ref{El}) for the longitudial part of the electric field
is negligible. The remaining terms in Eq. (\ref{El})
are local, so it makes sense to consider electric field of {\it one}
charge. From Eqs. (\ref{Ez}) and (\ref{El})
it is clear that this field is symmetric under the spatial rotations
on the brane. The field $E_z$ vanishes as $|z| \to \infty$, so 
eq.~(\ref{8*}) in its integral form becomes the $D$-dimensional 
Gauss' law,
\be
\int~d \sigma^i~E_i^{(l)} = e^2 q
\label{9+}
\ee
for the integrated longitudinal field
\[
   E_i^{(l)} = \int_{-\infty}^{+ \infty}~dz~ E_i(z, {\bf x})
\]
In eq.~(\ref{9+}), integration is performed over a 
$(D-2)$-dimensional sphere of radius $L$. Now, the longitudinal field
also vanishes at large $|z|$, so eq.~(\ref{9+})
implies 
\be
\label{long}
E^{(l)}\propto {1\over L^{D-2}}\;,
\ee
%where $E_{l}$ is an average longtudinal electric field.
%, defined as
%\be
%\label{average}
%E_{la}={1\over z_0}\int_{-\infty}^{\infty}~dz~E_l\;.
%\ee
%As it follows from Eqs.~(\ref{Ez}) and (\ref{El}) both $E_z$ and $E_l$
%exponentially decays in the bulk, hence the integral in
%Eq.~(\ref{average}) is saturated by the region $|z|\lesssim
%z_0$. Assuming that the $L$ dependence of $E_l$ is the same inside
%this interval we see 
This suggests that the interaction strength between the two 
charges indeed follows the
$D$-dimensional Coulomb law.

\subsection{Zero mode}
In order to substantiate the above semi-quantitative argument let us
study the spectrum of the longitudial operator (\ref{longD}).
This operator is Hermitean with the measure $\Lambda^{-2}(z)$, and is
positive definite.
The existence of the long-range field (\ref{long}) implies 
that the operator (\ref{longD})
has nearly zero mode with mass  
\be
\label{lightcond}
m_{light}\lesssim L_c^{-1}(0). 
\ee
To see this explicitly let us make use of the inequality
\be
\label{pauli}
\int {dz\over \Lambda^2(z)}\psi^*{\cal D}_L\psi 
>m_{light}^2\int {dz\over \Lambda^2(z)}\psi^*\psi\;,
\ee
where the weight is chosen according to the Hermiticity property of
${\cal D}_L$. Here $\psi(z)$ is an 
arbitrary continuous function vanishing at 
infinity. Let us consider the trial
function $\psi$ of the following form
\be
\psi(z)=\exp{\l-{1\over e}\int_0^zdz\bar{\Lambda}(z)\r}\;,
\ee
where the function $\bar{\Lambda}(z)$ is defined as follows,
\be
\bar{\Lambda}(z)=\mbox{sgn}(z)\Lambda(z)\;.
\ee
With this trial function, the estimate for the mass of the
lightest mode is
\be
\label{bound}
m_{light}^2\sim{2\int\psi^2dz\over e^2\int{dz\over\Lambda^2(z)}\psi^2}\;.
\ee
Now it is clear, that if $\Lambda(0)$ tends to zero, then the mass of
the lightest mode also vanishes. 
If one introduces parameters $z_0$ and $k^{-1}$ which
determine the widths of the regions near $z=0$ where $\psi(z)\sim 1$
and $\Lambda(z)\sim\Lambda(0)$, respectively, 
then from (\ref{bound}) one has the
following estimate (assuming $k>z_0^{-1}$)
\be
m_{light}^2\sim {1\over e^2}\Lambda^2(0)kz_0
\label{30}
\ee
from which it follows that the relation (\ref{lightcond}) is valid.

To illustrate the above general reasoning let us consider a specific 
choice of $\Lambda(z)$ in more detail. 
Namely, we chose $\Lambda(z)$  in the
form
\be
\label{choice}
\Lambda(z)=\Lambda_c\e^{k\l|z|-z_0\r}\;.
\ee
We  will assume the following relation between various
dimensionful parameters,
\be
\Lambda(0)=\Lambda_c\e^{-kz_0}\ll \l z_0^{-1}, {\Lambda_c\over e}\r\ll
k\ll \mu
\ee
To take the limit of small $\Lambda (0)$, we keep $\Lambda_c$ and
$z_0$ fixed, and take $k$ large. Then $z_0$ and $k$ are essentially 
the same parameters that enter our qualitative estimate,
eq.~(\ref{30}).
The eigenvalue equation for the operator (\ref{longD}) has the
following form at $z>0$,
\be
\label{eigeq}
-\psi''+2k\psi'+\l{\Lambda_c\over e}\r^2\e^{2k\l z-z_0\r}\psi=p^2\psi
\ee
%\be
%\label{bcond}
%\psi'(0)=0\;.
%\ee
The general bounded at  infinity solution to Eq. (\ref{eigeq}) is
\be
\label{solution}
\psi(z)=N\e^{kz}K_{\nu}\l{\Lambda_c\over ek}\e^{k(z-z_0)}\r\;,
\ee
where $N$ is a normalization factor and the order $\nu$ of the
modified Bessel function $K_{\nu}$ is equal to
\[
\nu=\sqrt{1-{p^2\over k^2}}
\]
Let us first consider symmetric eigenfunctions. They obey
\be
\label{bcond}
\psi'(0)=0\;.
\ee
which gives
\be
\label{bessels}
(1+\nu)K_{\nu}(\xi_0)-\xi_0 K_{\nu+1}(\xi_0)=0\;,
\ee
with
\be
\xi_0={\Lambda_c\over ek}\e^{-kz_0}={\Lambda(0)\over ek}\ll 1 \;.
\ee
Equation (\ref{bessels}) determines the eigenvalues $p^2$. The lowest
one is
\be
\label{estimate}
m^2_{light}=
2\l{\Lambda(0)\over e}\r^2\l k z_0-\ln\l{\Lambda_c \over ek }\r\r\;,
\ee
whereas higher eigenvalues are separated by the gap of order $k$ (in
fact, in the limit $\xi_0\to 0$, the next-to-lowest eigenvalue tends
precisely to $k$).
The inequality (\ref{lightcond}) is satisfied
and the light mode behaves as massless
at distances of interest, $L< L_c(0)$. Clearly, in the limit
$\Lambda(0)\to 0$, the light mode becomes exactly massless. It is this
massless mode that mediates $D$-dimensional Coulomb interaction
between charges on the brane. 
Note that the result (\ref{estimate}) is in agreement with our 
previous estimate (\ref{30}).

The anti-symmetric eigenfunctions obey
\[
    \psi (0) =0
\]
According to eq.~(\ref{solution}), this is possible only for 
$p > k$, when $\nu$ is imaginary. So, all eigenvalues  but one 
are large, and the light mode is the only one relevant at 
large distances.
%\begin{figure}[tb]
%\begin{center}
%\begin{picture}(30,70)(165,0)
%\put(100,0){\makebox(0,0)[lb]{a)}}
%\put(300,0){\makebox(0,0)[lb]{b)}}
%\psfig{file=fig1.ps,width=6.7cm,height=6.7cm}
%\psfig{file=fig2.ps,width=6.7cm,height=7.6cm}
%\end{picture}
%\end{center}
%\caption{a) L.h.s. of the Eq. (\ref{bessels}) for different values of
%the parameter $\xi_0$.\newline
%b) Numerical (solid line) and approximate analytical (dashed line)
%value of the mass $m_{light}^2$ of the light mode}
%\label{move}
%\end{figure}
%It worth noting, that the presence of light mode found above implies
%the validity of the assumption made after Eq.~(\ref{average}) that the
%$L$-dependence of the long-range electric field does not change with
%$z$. Indeed, far from the sources this field is given by
%\[
%E_l(z,x)\propto E_{light}(z)/L^{D-3}\;,
%\]
%where $E_{light}$ is the light mode.

\subsection{Charge universality}
Let us now discuss the charge universality in this
set up. 
To this end, let us consider two point-like opposite charges 
displaced from the brane to the points $z_{+}$ and $z_{-}$,
respectively. Clearly, the general case of a continuous charge 
distribution in $z$-direction can be 
straightforwardly obtained from this one.

Naively, one may expect that the following
two effects take place. First, 
the confinement length between two charges may appear to be 
determined now not by $\Lambda(0)$ as in Eq. (\ref{confscale}), but by
$\Lambda(z_c)$ where $z_c$ takes 
some intermediate value between $z_+$ and $z_-$. Second,
in
the distance interval where confinement between the
two charges has not
set in yet, the long range force is transmitted by the
light mode. Then the general argument, described in Introduction,
suggests that this force depends on the overlaps between the charge
distributions and wave function of the light mode. In other words,
one might expect that the force between the two charges is
proportional to $\psi_0 (z_+)\cdot \psi_0 (z_-)$, where $\psi_0 (z)$ 
is the light mode. This would mean the lack of charge universality.
However, we will see that neither of these two effects actually occurs.

As before, let us take the distance between the charges
along the brane, $L$, to be much smaller than the confinement length
on the brane, $L_c (0)$. The key point is that outside 
an $L$-independent neighborhood of each of the charges, the minimal 
string {\it lays  on the brane}: clearly, any other string configuration 
has higher energy and action. This configuraion is illustrated in
Figure 1, where the electric field lines are also shown.

\begin{figure}[tb]
\begin{center}
\begin{picture}(70,70)(165,0)
\epsfig{file=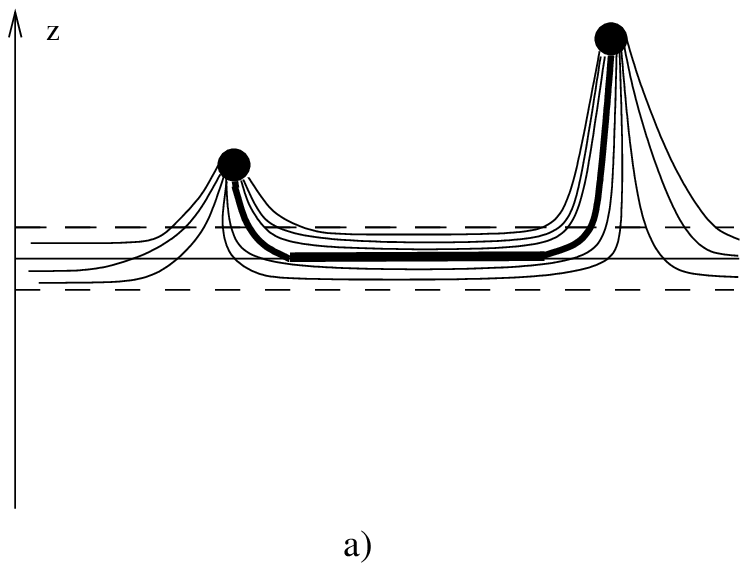,height=6.cm,width=7.cm}
\epsfig{file=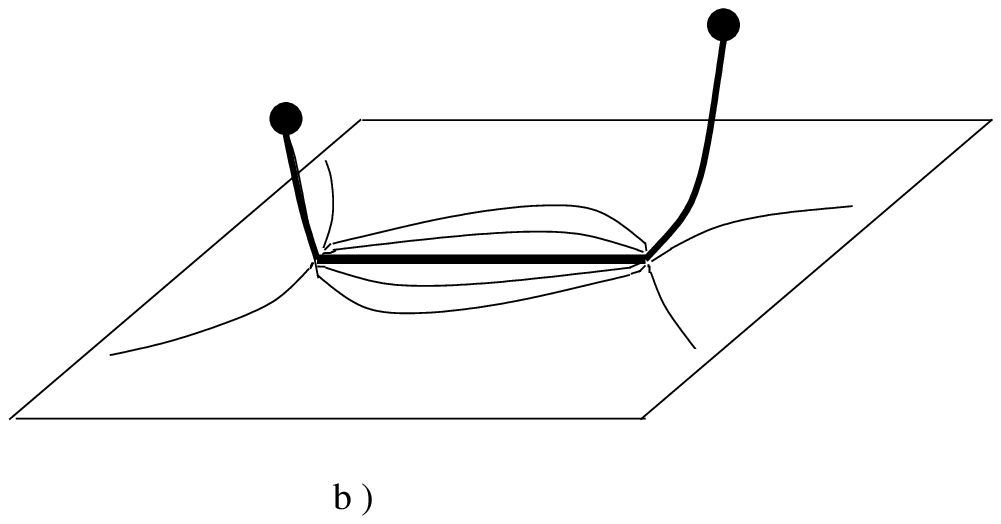,height=6.cm,width=7.cm}
\end{picture}
\end{center}
\caption{A string (thick line) and electric field lines for two
charges placed in the bulk: a) side view; b) overall picture (electric field
lines are shown on the brane only). }
%b) Power lines of the electric field of the ``free quark''}
\end{figure}

Then, as in the case of point charges located on the brane, there
are two different components of the electric field. The first one
is non-local and originates from the long string running along the 
brane.
It is this component that is responsible for linear potential between 
the charges, so confinement scale is $L_c(0)$, as before.
The second component is (almost) local and comes 
from the regions
near the charges; it includes, in particular, the contributions 
of the parts of the string which connect the charges to the brane.
The non-local component is suppressed
by the small value of the confinement scale on the brane, $\Lambda(0)$
(cf. Eqs.~(\ref{wzmu}) and (\ref{wmunu})), and is negligible 
at the distances between charges smaller than $L_c(0)$, which is the
case we consider. For the local component it makes sense to consider
electric fields of 
each of the charges separately.
Sufficiently far away from the charges, 
the long range component of
this field
is due to the light mode and has the form
\be
\label{longrange}
E_i = \hat{{\bf x}}_i \frac{a}{\Omega_{D-2}}{\psi_0(z)\over L^{D-2}}\;,
\ee
where $L$ is the distance to the charge, $\Omega_{D-2}$ is
the area of unit $(D-2)$-dimensional sphere and $\hat{{\bf x}}$ is 
unit radius-vector.
The coefficient $a$ is determined by the
Gauss' law (\ref{9+}),
\be
\label{a}
a={qe^2\over\int dz\, \psi_0(z)}\;.
\ee
We see that this coefficient is independent of the 
position of the charge in transverse direction, 
and is determined by the value of charge $q$ only.
Hence, charge universality holds in our model.

To find the effective charge $q_{eff}$ 
in the low-energy effective theory on the brane, let us calculate the
interaction energy of the two charges. As in the usual
electrodynamics, the Gaussian structure of the action (\ref{action})
implies that this energy may be calculated as an energy of the
two-form field $\omega_{0i}=iE_i$  produced 
by the charges. 
The contribution into the energy integral that dominates at large $L$, 
comes from the region where both $|{\bf x} - {\bf x}_+|$ and
 $|{\bf x} - {\bf x}_-|$ are of order $L$, where ${\bf x}_{\pm}$
are the longitudinal coordinates of the charges. The regions near 
the charges contribute into higher multipoles only\footnote{
It is interesting to note that higher multipoles are {\it not} universal,
as they depend on $z_{\pm}$.}.

Using
Eq.~(\ref{longrange}), one obtains for the interaction energy
\[
V(L) = -\frac{a^2}{ 4 \Omega_{D-2}^2}
\left(\int \frac{d^{D-1} {\bf x}}{|{\bf x} - {\bf x}_+|^{D-2} 
|{\bf x} - {\bf x}_-|^{D-2}} \right) \int~dz 
\left[ {1\over
12\Lambda^2}\l \d_z \psi_0 \r^2+{1\over 4e^2} \psi_0^2\right]\;.
\]
The integral over longitudinal coordinates ${\bf x}$ gives the
$D$-dimensional 
Coulomb behavior, $V (L) \propto 1/L^{D-3}$, precisely in the same 
way as in the usual electrodynamics.
Evaluating the integral over $z$ by parts
 and using the fact that $\psi_0(z)$ is an
eigenmode of the operator ${\cal D}_L$ with the eigenvalue
$m_{light}^2$ we obtain
\[
q_{eff}^2=a^2m_{light}^2\int dz\,{\psi_0^2(z)\over \Lambda^2(z)}
\]
or, making use of Eq.~(\ref{a}),
\[
q_{eff}^2=q^2e^4m_{light}^2{\int dz{\psi_0^2(z)\over
\Lambda(z)^2}\over\l\int dz\,\psi_0(z)\r^2}\;.
\]
Now, recalling the estimate (\ref{bound}) for $m_{light}$ we find
that $q_{eff}$ remains finite in the limit $\Lambda(0)\to 0$. 
Under the same conditions that lead to eq.(\ref{30}), 
the
estimate for the effective charge is
\[
q_{eff}^2\sim {q^2\over z_0}.
\]
To summarize, in the low energy effective theory, the charges
are finite and universal.

% \begin{figure}[tb]
%\label{surface}
%\begin{center}
%\begin{picture}(70,70)(165,0)
%\epsfig{file=fig3.eps,height=6.cm}
%\epsfig{file=fig4.eps}
%\end{picture}
%\end{center}
%\caption{a) Power lines of the electric field for two charge
%distributions of finite size; 
%b) Power lines of the electric field of the ``free quark''}
%\end{figure}

\subsection{Discussion}

In our model, the $d$-dimensional Coulomb law between charges
placed on or near the brane comes together with 
 the zero mode of the gauge field. This is in accord with the
general rule of the existence of a massless particle, photon, 
travelling along the brane, in theories exhibiting
the Coulomb law on the brane. Another important ingredient are
thin strings between the charges. These strings are
invisible in the low energy effective theory (for $\Lambda(0) \to 0$),
as they lay on the brane, but play a key role in ensuring
charge universality.
For charges displaced from the brane (or non-trivial charge
distributions in the transverse direction), the   
short string connecting the charge to the brane should 
contribute to the self-energy of the charge. There may be other 
low-energy effects of these short strings, such as polarizability of
charged particles whose wave functions have finite spread in the
transverse direction.

The model discussed in this section has exotic modifications.
For instance one may consider several branes parallel to each other as
in Ref.~\cite{Arkani-Hamed:1999ww}.
Another modification  with potentially interesting
phenomenological consequences is as follows. 
Suppose that the confinement scale on our
brane is finite, as it happens in QCD, and that
$\Lambda(z)$ has a local minimum on our brane. Suppose further
that $\Lambda(z)$ has a global minimum on some other brane or in the
bulk, and that $\Lambda = 0$ there. Then there should exist 
"free quarks" whose electric field is shown in Figure 2.
\begin{figure}[tb]
\begin{center}
%\begin{picture}(70,70)(,0)
%\epsfig{file=rub.eps,height=6.cm}
\epsfig{file=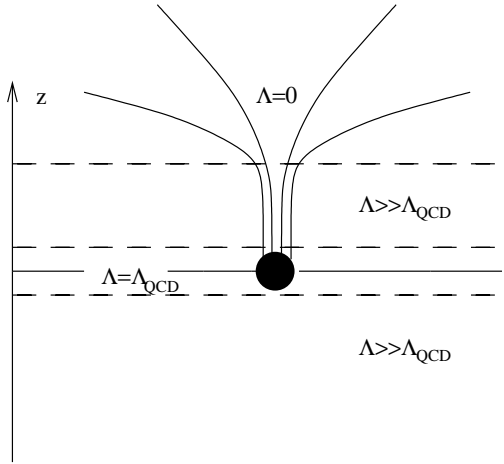}
%\end{picture}
\end{center}
\caption{Color-electric field lines of a ``free quark''}
\end{figure}
These objects will be colorless, but otherwise have quantum 
numbers of a quark. Their masses will be determined by the properties of
the flux tube, extending into extra dimensions and connecting the
free quarks to the region with $\Lambda =0$; this mass may naturally be
quite large.

\section{Brane-induced localization}

Let us now turn to another proposal \cite{Dvali:2001rx}. Consider 
$(4+d)$-dimensional flat space-time with coordinates $x^{\mu}$
($\mu = 0,1,2,3$) and $z^{\alpha}$ ($\alpha = 1, \dots , d$), and
assume that there is a brane at $z^{\alpha} = 0$. It has been 
proposed to choose the action for the gauge field in the following 
form,
\be
  S = \frac{1}{g_{(4+d)}^2}
\int~d^4 x~d^d z~\left( \frac{1}{4} F_{ab}^2 + \frac{1}{4}
C^2 \delta^d (z) F_{\mu \nu}^2 \right)
\label{19*}
\ee
where $F_{ab}$ is the $(4+d)$-dimensional field strength,
$F_{\mu \nu}$ are its four-dimensional components, $C$ is some 
constant of dimension $M^{-d/2}$; we work in Euclidean space-time for
convenience. The second term in eq.~(\ref{19*}) concentrates on the 
brane; the discussion of how the term of this type may be generated
is given in Refs.~\cite{Dvali:2000hr,Dvali:2001xg,Dvali:2001rx}.

\subsection{More than two extra dimensions}

To be specific, let us first consider the case
$ d>2$.
A heuristic argument suggesting that charges residing on the brane 
experience the four-dimensional Coulomb law is as follows (cf.
Ref.~\cite{Dvali:2001xg}). Omitting indices, and ignoring 
complications due to gauge fixing, one writes the equation for the 
propagator,
\be
\left( \Box^{(4+d)} + C^2 \delta^{d} (z) \Box^{(4)}\right)
G(x,x'; z,z') = - \delta^4 (x-x') \delta^d (z-z')
\label{19a*}
\ee
Placing the source on the brane, {\it i.e.} setting
$z'=0$, and using the four-dimensional momentum
representation, one has 
\be
\left( \Box^{(d)} - p^2 - C^2 p^2 \delta^{d} (z) \right)
G(p; z) = - \delta^d (z)
\label{gpz}
\ee
where $p_{\mu}$ is the four-momentum.
A formal solution to this equation is
\be
  G(p;z) = \frac{D(p;z)}{1 + C^2 p^2 D(p;0)}
\label{ii*}
\ee
where $D(p;z)$ is the free $(4+d)$-dimensional propagator 
in a space-time without brane, which obeys
\be
\left( \Box^{(d)} - p^2 \right) D(p;z) =  - \delta^d (z)
\ee
Now, at $d>2$, the free propagator $D(p;z)$ is finite at 
finite $z^{\alpha}$ and
diverges as $ z^{\alpha} \to 0$. So, one argues that 
$G(p;z) = 0$ at $ z^{\alpha} \neq 0$, and
\be
  G(p;0) = \frac{1}{C^2 p^2} \;\;\; \mbox{at} \;  z^{\alpha} = 0
\label{19b*}
\ee
This quantity is proportional to the
 four-dimensional propagator, so one argues 
that the charges on the brane experience the four-dimensional 
Coulomb law. The effective four-dimensional 
gauge coupling would then be equal to
\be
  g_{(4)} = \frac{g_{(4+d)}}{C}
\label{19b+}
\ee
where $g_{(4+d)}$ is the gauge coupling in the original theory.

Of course, this argument is far from being rigorous
\cite{Carena:2001xy}. The right hand side of eq.~(\ref{ii*})
does not exist (even in distributional sense), as $D(p;0)$ is
infinite. The $\delta$-function in the action (\ref{19*}) has to be
regularized. One way to do this is to smear this $\delta$-function
over a spherical region of small size $z_0$; it is precisely this
regularization that we use in this paper.

As pointed out in Introduction, it is natural to suspect that this
proposal has problems with charge universality.
The above arguments imply that only
that part of the charge which is contained in the region
$|z| < z_0$ will participate in the four-dimensional Coulomb 
law  (hereafter
$|z| \equiv \sqrt{z^{\alpha}z^{\alpha}}$). 
The zero modes of matter fields may, however, spread over larger
distances in transverse dimensions, so the four-dimensional charges
of matter fields may depend on the shapes of their zero modes.

To see what the above set up actually corresponds to, let us 
consider the  action
\be
  S = 
\frac{1}{g_{(4+d)}^2}
\int~d^4x~d^d z~\left( \frac{1}{4} F_{ab}^2 + \frac{1}{4}
f^2(z)F_{\mu \nu}^2 \right)
\label{19c*}
\ee
where $f(z)$ is a step function equal to a constant $f_0$ at
$|z| < z_0$ and zero at $|z| > z_0$. 
Omitting factors
of order unity, we have a relation between parameters entering 
eqs.~(\ref{19*}) and (\ref{19c*}),
\be
  z_0^{\frac{d}{2}} f_0 \sim C
\label{19c+}
\ee
Clearly, $f_0$ is large for small $z_0$.

Let us assume for simplicity that matter currents have  
components, transverse to the brane, equal to zero. 
In this case we can set the transverse components of
the gauge field equal to zero, $A_{\alpha} = 0$, and write the action
explicitly as follows,
\be
   S = 
\frac{1}{g_{(4+d)}^2}
\int~d^4x~d^dz~ \left[ \frac{1}{2} (\partial_{z^{\alpha}} 
A_{\mu})^2
+ \frac{1}{4} \hat{f}^2 (z) F_{\mu\nu}^2 \right]
\label{20+} 
\ee
where $\hat{f}^2 (z) = 1 + f^2 (z)$. Let us consider the region
$|z| < z_0 $, in which  
$\hat{f}^2  = \mbox{const} = 1 + f^2 \equiv \hat{f}_0^2$.
Since $f_0 \gg 1$, we will not distinguish between the constants
$\hat{f}_0$ and $f_0$ in what follows. The propagator in this 
region is\footnote{In fact, the correct expression for the
propagator should be obtained by finding  the solutions
inside ($|z|<z_0$) and outside ($|z|>z_0$) 
the brane, and matching them
at $|z|=z_0$. This leads to corrections to the
expression~(\ref{20*}). For sources which are spherically symmetric 
in transverse dimensions, the propagator can be found in explicit 
form. In this way one can show that corrections to eq.~(\ref{20*})
are negligible at distances of interest.} (again omitting indices)
\be
  G(x,x'; z,z') = \int \frac{d^4p~d^dq}{(2\pi)^{4+d}}~
\frac{\mbox{e}^{iq(z-z')} \mbox{e}^{ip(x-x')}}{q^2 + f_0^2 p^2}
\label{20*}
\ee
Now, the typical transverse momenta are of order $q \sim 1/z_0$,
whereas the four-dimensional momenta are of order $p \sim 1/r$
where $r$ is the distance along the brane.
So, there is a characteristic distance scale
\be
   r_c = f_0 z_0 \sim \frac{C}{z_0^{\frac{d}{2} -1}}
\label{21*}
\ee
where we made use of eq.~(\ref{19c+}). This scale is 
large at small $z_0$ (recall that we consider the case $d>2$).
Below this scale, the transverse momentum $q$ in the denominator 
in eq.~(\ref{20*}) is negligible, and the propagator is
\be
   G(x,x'; z,z') = \frac{1}{f_0^2} D^{(4)}(x-x') \delta (z-z')
\label{22*}
\ee
where $D^{(4)}$ is the conventional massless four-dimensional 
propagator. This is how the expectation concerning
the four-dimensional Coulomb law \cite{Dvali:2001xg,Dvali:2001rx}
is confirmed. The propagator has four-dimensional behaviour
at $r \ll r_c$, while extra dimensions "open up" at $r \sim r_c$.

To see how the relation (\ref{19b+}) emerges,
let us consider two static charge distributions $\rho(z)$ and $\rho'(z)$
spreading over  the region $|z| < z_0$ and separated by a
distance $r \ll r_c$ along the brane. According to eq.~(\ref{22*}),
the potential between them is
\be
   V(r) = \frac{g^2_{(4+d)}}{4 \pi r} ~ \frac{1}{f_0^2}
\int~d^dz~\rho(z) \rho'(z)
\label{23*}
\ee
Now, $\rho \sim q/ z_0^d$, $\rho' \sim q'/ z_0^d$, where $q$ 
and $q'$ are the total charges,  so
\be
   V(r) \sim  \frac{g^2_{(4+d)}}{f_0^2 z_0^d}~ \frac{q q'}{4 \pi r}
\ee
Taking into account eq.~(\ref{19c+}), one indeed obtains the relation 
(\ref{19b+}).

The result (\ref{23*}) is alarming, however. Not only the charge 
universality does not hold, but also the interaction between
the sources is {\it ultra-local} in $z$. This calls for further 
analysis of this model.

Let us again consider the action
(\ref{20+}) inside the region $|z| < z_0$ and change variables 
from $z^{\alpha}$ to
\[
 y^{\alpha} = f_0 z^{\alpha}
\]
In terms of these variables, the action (\ref{20+}) becomes
\be
S = 
\frac{1}{g_{(4+d)}^2 f_0^{d-2}} 
\int~d^4x~d^dy~ [ (\partial_{y^{\alpha}} 
A_{\mu})^2
+ F_{\mu\nu}^2]
\label{fin}
\ee
This is the standard $(4+d)$-{\it dimensional} action for
($\mu$-components of) the gauge field. Now it is clear why
the interaction (\ref{23*}) is ultra-local: in terms of the new 
variables, the charge distributions are huge pancakes of the
transverse size
\[
  y_0 = f_0 z_0 \sim r_c
\]
whereas the gauge fields propagate with the speed of light in all
directions. At $r \ll r_c$, i.e., when the distance between the 
charged pancakes
is much smaller  than their size, the interaction occurs between the
pieces of pancakes sitting in front of each other, hence the
ultra-locality.

From the point of view of gauge fields themselves,
space-time inside the brane, $|y| < y_0$, is $(4+d)$-dimensional,
flat and has the transverse size of order $r_c$. This is unacceptable,
at least in the non-Abelian case. In the first place, there are 
many more degrees of freedom than in the four-dimensional theory.
Furthermore, according to eq.~(\ref{fin}),
the effective $(4+d)$-dimensional
self-coupling of the gauge fields is
\be
  g^{eff}_{(4+d)} = g_{(4+d)} f_0^{\frac{d}{2} -1} 
\sim g_{(4)} r_c^{\frac{d}{2}}
\label{j*}
\ee
where we made use of eqs.~(\ref{19c+}) and (\ref{21*}).
Thus, for $g_{(4)}$ roughly of the order of unity, the gauge theory
inside the brane becomes strongly coupled at distances
(in all directions) just below $r_c$. This situation cannot be treated
semiclassically, contrary to what has been implicitly
assumed throughout the whole discussion.

The discussion of this subsection applies, word for word, to 
brane-localized gravity in more than two extra dimensions, with the
obvious substitution
\begin{eqnarray}
g^2_{(4+d)} \; \; & \to & \;\; L^{2+d}_{(4+d)} \equiv \frac{1}{M^{2+d}} 
\nonumber \\
g^2_{(4)} \; \; & \to & \;\; L^{2}_{Pl} \equiv \frac{1}{M^{2}_{Pl}}
\end{eqnarray}
where $M$ is the gravity scale in the underlying $(4+d)$-dimensional 
theory and
$M_{Pl}$ is the four-dimensional Planck mass. Gravity inside the brane
ceases to be four-dimensional, and becomes strong at distances below
the length scale obtained from eq.~(\ref{j*}),
\[
         L_{(4+d)}^{eff} = (L_{Pl}^2 r_c^d)^\frac{1}{2+d}
\]
For $d=3$ and $r_c \sim 10$~kpc, this effective scale is of order
1~cm, which is unacceptably large. The scale $ L_{(4+d)}^{eff}$ is
even larger at $d>3$.

\subsection{One extra dimension}

The cases $d=1$ and $d=2$ are special.
Brane-induced localization in two extra dimensions has similar
problems as in the case $d>2$: at $d=2$, the free propagator $D(p;z)$
is again singular at $z=0$, so the discussion of the beginning of the
previous subsection applies to the case $d=2$ as well. The case $d=1$
is different. Let us consider the latter case in some detail.

Let us assume for simplicity that charge distributions are symmetric
under $z \to -z$, and consider the symmetric part of the 
propagator $G(x,x';z,z')$. At $z,z' > 0$, the solution to
eq.~(\ref{19a*}) with $d=1$ is (in four-dimensional momentum
representation) 
\be
  G(p;z,z') = \frac{1}{2(r_c p^2 +p)}
\left[ r_c p \left( \mbox{e}^{-p|z-z'|} - \mbox{e}^{-p(z+z')}\right)
+
\left( \mbox{e}^{-p|z-z'|} + \mbox{e}^{-p(z+z')}\right) \right]
\label{v2*}
\ee
where
\[
   r_c \equiv C^2 = \frac{g_{(5)}^2}{g_{(4)}^2}
\]
This propagator is finite at $z=0$ and/or $z'=0$, so one does
not need to regularize the $\delta$-function in the action
(\ref{19*}).
For sources on the brane, $z=z'=0$, the propagator (\ref{v2*})
agrees with the expression given in Ref.~\cite{Dvali:2000hr},
\be
   G(p;0) = \frac{1}{r_c p^2 + p}
\label{v22*}
\ee
Charges placed on the brane at distance $r \ll r_c$ apart, experience
the four-dimensional Coulomb law.

Charge distributions of width $z_0$ in transverse direction also
experience the four-dimensional Coulomb law, provided the first term
in square brackets in eq.~(\ref{v2*}) is small compared to the second
term. The latter requirement gives
\[
        \frac{r_c z_0}{r^2} \ll 1
\]
This automatically implies that $z_0 p \sim z_0/r \ll 1$, so in this
regime the propagator is independent of $z$ and $z'$ and has the
form (\ref{v22*}). This means that 
the interaction between the charges involves
integrals
$\int~dz~\rho(z,x)$, so the charge universality holds automatically in
the effective four-dimensional theory. 

In non-Abelian case, gauge field self-interaction in the bulk occurs
with the coupling
\[
          g_{(5)} = \sqrt{r_c} g_{(4)} 
\]
With $g_{(4)}$ roughly of order one,  the gauge theory in the bulk is
strongly coupled at all distances of interest, so the brane-induced
mechanism is not suitable for localizing non-Abelian gauge fields.

Translating to gravity, we find that the effective four-dimensional
theory obeys the equivalence principle, and the four-dimensional
Newton's law between smooth sources is valid at distances
\be
   r_c \gg r \gg \sqrt{r_c z_0}
\label{v23*}
\ee
where
\[
          r_c =\frac{L_{(5)}^{3}}{L_{Pl}^2}
\]
The interval (\ref{v23*}) is large enough if $z_0$
is sufficiently small. Requiring $r_c > 10$~kpc and 
$ \sqrt{r_c z_0} > 0.1$~mm, we find
\be
       z_0 < 10^{-26}~\mbox{cm}
\label{finn}
\ee

There are two problems with this scenario. One is the
scalar-tensor structure of the four-dimensional graviton propagator
\cite{Dvali:2000hr}. Another stems from the argument of
Ref.~\cite{Carena:2001xy} that one can trust the calculations
leading to eq.~(\ref{v2*}) only if the transverse distances are larger
than $L_{(5)}$. For $r_c \sim 10$~kpc one has
$L_{(5)} \sim 10^{-15}$~cm, and from (\ref{finn})
we find that $z_0 \ll L_{(5)}$, in conflict
with
the latter argument. It remains to be understood how serious these two
problems are; one approach to get around at least some of these
problems is suggested in Ref.~\cite{Dvali:2001gm}.

We are indebted to C.Deffayet, J. Lykken,
M.Shaposhnikov, S.Sibiryakov,
P.Tinyakov, S.Troitsky, and M.Voloshin for helpful discussions. This work
was supported in part by RFBR grant 99-01-18410, by the Council for
Presidential Grants and State Support of Leading Scientific Schools,
grant 00-15-96626, by Swiss Science
Foundation grant 7SUPJ062239 and by CRDF grant (award RP1-2103).

\end{document}